\documentclass[aps,amssymb,amsmath,preprint,showpacs]{revtex4}
\begin{document}
\preprint{}
\title{Analytic calculation of energy transfer and heat flux in a 
one-dimensional system}
\author{V. Balakrishnan}
\affiliation
{Department of Physics,
Indian Institute of Technology-Madras, Chennai 600 036, India}
\email{vbalki@physics.iitm.ac.in}
\author{C. Van den Broeck}
\affiliation{Limburgs Universitair Centrum, B-3590 Diepenbeek, Belgium}
\email{christian.vandenbroeck@luc.ac.be}
\date{\today}

\begin{abstract}
In the context of the problem of heat conduction in one-dimensional
systems, we present an analytical calculation of the instantaneous 
energy transfer across a tagged particle in a
one-dimensional gas of equal-mass, hard-point particles. 
From this, we obtain  
a formula for the steady-state energy flux, 
and identify and separate  
the mechanical work and heat conduction contributions  
to it. The nature of the Fourier law for the model, and 
the nonlinear dependence of the rate of mechanical work 
on the stationary drift velocity of the tagged particle, are 
analyzed and elucidated. 
\end{abstract}
\maketitle

\section{Introduction}
\label{intro}

Heat conduction in one-dimensional systems has evoked a considerable
amount of interest in recent years. 
In particular, the validity or otherwise 
of the  Fourier law of heat conduction
in such systems is a non-trivial question, and hence one that has been the
subject of lively discussion\cite{dhar}. 
A comprehensive account of the current status of the problem is
provided in Ref. \cite{lepr}. 

Most of the results known in this regard 
are based on numerical studies of lattice models.
An alternative and useful line of development is the analytical study of a
model system, albeit a simplified one, which we call the Jepsen gas: 
a system of $N$ identical classical  point-particles of mass $m$ moving on
a line and undergoing perfectly elastic collisions when neighboring
particles meet\cite{jeps}. This system is a special case 
of the more general
model investigated in the context of the so-called 
adiabatic piston problem\cite{lieb}, in which a central heavy particle of mass
$M$ is in a gas of particles of mass $m$ on its left and right. The
problem then is to analyze the dynamics of the central particle in the
thermodynamic limit in which the one-dimensional 
gases to its left and right are in thermal equilibrium at specified
temperatures and densities. The Jepsen gas is a {\em singular} limiting case
of the unequal mass situation. The latter is not integrable, in marked
contrast to the 
case $M = m$, which is integrable. Notwithstanding this simplification,    
non-trivial irreversible behavior is observed  in the Jepsen gas 
in the limit  $N \rightarrow \infty$
when appropriate averages over initial conditions
are performed. In the initial studies of this system\cite{jeps,lebo},  
several quantities of interest 
such as the statistics of the displacement and velocity of one of the
particles, which we shall refer to as the central or tagged particle (or
``piston'', as it is the counterpart of the adiabatic piston in the
model at hand),   
have been calculated exactly
by performing an average over equilibrium
initial conditions for the rest of the particles. 
In particular, it can be shown that the motion of this
tagged particle becomes diffusive asymptotically, 
i.e., converges to Brownian motion. 

Recently, the model has been revisited and 
studied in greater detail\cite{pias,bala1},
one of the motivating factors being its relationship to the adiabatic piston
problem.  The non-equilibrium situation implied by
different velocity distributions for the gases to the left and right
of the tagged particle has been analyzed. In general, this
particle acquires, in the thermodynamic limit, a
systematic drift velocity over and above its diffusive motion. A 
notable feature is that this  drift velocity is exclusively
fluctuation-induced\cite{pias}. In the special case when the velocity
distributions of the gases on the two sides are Maxwellian, so that
the pressure of each gas can be identified with $k_{B}$ times the 
product of its (linear) number density and temperature, an interesting
feature emerges: even when the pressures of the two gases are equal,
the drift velocity of the tagged particle does not vanish. Rather, it is
directed from the lower temperature (higher density) side to the 
higher temperature (lower density) side. The drift velocity vanishes when
the product of the number density and the {\em square root} of the
temperature is the same on the two sides of the tagged particle, a
condition already recognized in Ref. \cite{jeps}. 

In the context of heat conduction in one dimension, the question that 
arises naturally is whether the heat flux can be calculated for the
Jepsen gas. We show in this paper that this problem can be solved
analytically: it is possible to calculate 
exactly the amount of energy that
is transfered through the tagged particle or piston. At first, 
it would appear that
the issue is a trivial one in the same way as it is in a 
linear chain of harmonic
oscillators. In the latter system, the energy just travels
ballistically, being carried by phonon modes which do not interact
with  
each other.
In a similar fashion, the kinetic energy carried by any particle in the
Jepsen gas is just transfered upon collision to the next particle, and hence
moves ballistically along the line. As a result, a Fourier law, which predicts
the diffusive spreading of thermal energy, does not appear to be valid.
We will show by an exact and explicit calculation that this
``hand-waving'' argument, while formally correct, 
is nevertheless misleading in the sense that a
non-trivial energy flux obtains in the model. A closed expression can
be derived for this quantity, comprising two components. One of these involves 
the asymptotic drift velocity of the tagged particle,
while the other is present even when the drift velocity vanishes. This
permits the identification, and hence a natural separation, of the ``mechanical
work'' and ``heat'' contributions, respectively, to the energy
transfer. In particular, for initial conditions corresponding
to thermal equilibrium of the gases to the left and right of the 
tagged particle
(with densities $n^-$ and
$n^+$ and temperatures $T^-$ and  $T^+$, respectively) such that  
$n^-\sqrt{T^-}=n^+\sqrt{T^+}$, so that the drift velocity is zero, 
the asymptotic 
steady-state energy flux through the tagged particle is shown to 
converge to a constant
value.  For small values of $(T^+ - T^-)\,,$ this quantity is linearly 
proportional to the temperature difference itself, which 
can be interpreted as a manifestation of the Fourier law.
Moreover, the ``mechanical
work'' contribution itself is shown to have  
a part that is linear in the drift velocity, as might be expected, as
well as a {\em nonlinear} part that starts (for small drift
velocities) with the fourth power of the drift velocity.

We reiterate the following point. The Jepsen gas is admittedly a
simplified special case of the more ``realistic'' models of
one-dimensional transport that have been the subject of much
attention. Nevertheless, it is the fact that analytical (and hence
unambiguous) results can be obtained for this model 
that makes it worth studying,
because these results help shed light on
several of the essential issues involved in energy
transport in one dimension. 

The plan of the rest of this paper is as follows. In the next
section, we introduce the notation and summarize the salient results
pertaining to the Jepsen gas for the purpose at hand. In
Sec. \ref{exactexpression} (and the Appendix), an exact formula is derived 
for the energy transfer across the piston at any instant of
time. This leads to a formula for the total energy flux in the stationary
state, which is obtained and analyzed in Sec. \ref{fluxformula}. 
In Sec. \ref{statflux}, the
contribution to the energy flux coming from the stationary heat flux
is identified, and the nature of the Fourier law for the model is
clarified. The rate of mechanical work is also deduced, and its
nonlinear dependence on the stationary drift velocity elucidated. 
Section \ref{conclusion} contains a few concluding remarks.

\section{Notation and recapitulation}
\label{notation}

It is helpful to recapitulate in brief the relevant features of the
model, in the notation used in earlier
work\cite{pias,bala1}. The tagged particle, 
located at $X = 0$ at $t=0$ with an initial velocity $V_{0}$,
separates a gas of $N^{-}$ particles in the interval $[-L,0)$ on its left
from a gas of $N^{+}$ particles to its right, in $(0,L]$.  Their initial
positions $X_{j}$ (where $-N^{-} \leq j \leq -1$ 
for the gas on the left, and $1
\leq j \leq N^{+}$ for the gas on the right) are independently and uniformly
distributed in the corresponding intervals.  Their initial velocities
$V_j$ are drawn from normalized distributions $\phi^{-}(V)$ and $\phi^{+}(V)$,
respectively.  To avoid unnecessary complications, 
we shall assume that these are symmetric distributions, i.e., $\phi^{\pm}(V) =
\phi^{\pm}(-V)$.  It can be shown that the system has a thermodynamic
limit in which $N^{+} \rightarrow \infty$, $L\rightarrow \infty$ with finite 
densities $\lim
N^{\pm}/L = n^{\pm}$, provided only that 
the mean speeds $\langle |V|\rangle^{\pm} =
\int_{-\infty}^{\infty} dV \, V\,\phi^{\pm} (V)$ are finite.  
Note that $\phi^\pm$ need not be Maxwellian distributions, although
this is the case of direct interest in the present context of heat
conduction. As all
the particles (including the tagged particle) have equal masses, they merely
exchange their identities on their original linear trajectories $X_{j}(t)
\equiv X_{j} + V_{j}\,t$ in each collision.  This is what enables one to derive
exact results for the system.  Among other quantities, the exact
one-particle distribution function of the tagged particle, 
$P(X,V,t \arrowvert 0,V_{0})$,
can be found.  The complicated stochastic 
motion of the tagged particle (henceforth termed the ``piston'') 
can be interpreted\cite{bala1} as being driven
by two independent Poisson processes (corresponding to collisions from its
left and right, respectively) with {\it state-} 
and {\it time-dependent} intensities
$n^{-}\alpha (X/t)$ and $n^{+} \beta (X/t)$, where 
\begin{equation} 
\alpha (W) =
\int_{W}^{\infty} dV \,(V-W) \,\phi^{-} (V)\,,\,\,  \beta (W) =
\int_{-\infty}^{W} dV \,(W-V) \,\phi^{+} (V)\,. 
\label{alphabeta1}
\end{equation} 
The system does not equilibrate in the conventional sense
of the term: the initial set of trajectories persists
for all $t$.  The piston  
acquires an asymptotic mean drift velocity
$\overline{W}$ given by the unique solution of the 
implicit equation \cite{pias}
\begin{equation} 
n^{-}\alpha(\overline{W}) = n^{+}\beta (\overline{W})\,.
\label{Wbar1} 
\end{equation} 
The asymptotic velocity
distribution of the piston is a superposition of $\phi^{+}(V)$ and
$\phi^{-}(V)$, and is given by 
\begin{equation} 
P_{\rm st}(V) = \left[n^{-} \phi^{-}(V)
\theta (V-\overline{W}) 
+ n^{+}\phi^{+}(V) \theta(\overline{W} - V)\right]/\,\Xi
(\overline{W})\,,
\label{Pstationary1} 
\end{equation} 
where 
\begin{equation}
\Xi(\overline{W}) = n^{-} \,|\alpha'(\overline{W})| + n^{+} \,\beta'
(\overline{W}) = n^{-} \int_{\overline{W}}^{\infty} dV \,\phi^{-} (V) + n^{+}
\int_{-\infty}^{\overline{W}} dV \,\phi^{+} (V)  
\label{chiwbar}
\end{equation} 
is the normalization factor.  Only in the special case of
the {\em homogeneous system}, 
defined by $n^{-} = n^{+} = n$ and $\phi^{+}(V) =
\phi^{-}(V) = \phi(V)$, can one have stationarity and an 
approach to equilibrium, in the sense
that $\overline{W} =0$ and $P_{\rm st}(V) \equiv \phi(V)$.  Note that
$\overline{W}$ may vanish even in the {\em inhomogeneous} system, if the
condition $n^{-}\int_{0}^{\infty} dV\, V \,\phi^{-}(V) = n^{+}
\int_{0}^{\infty} dV \,V \,\phi^{+} (V)$ (or 
$n^{-}\langle |V|\rangle^{-} = n^{+} \langle |V| \rangle^{+}$) 
happens to be satisfied: that is, 
the {\em mean rates} at which the piston suffers
collisions from its left and right, respectively, are equal.

We point out at this juncture that the 
stationary distribution $P_{\rm st}(V)$ in 
Eq. (\ref{Pstationary1}) may in fact be
anticipated on physical grounds, as follows. In the equal-mass case under
consideration, we may regard the particles as simply ``passing through''
each other in each collision, because they are identical particles.  In
the stationary state, after the memory of the initial velocity of the
piston has been lost, it will only encounter  
such particles via collisions from its left, as have velocities {\em
greater} than its asymptotic mean drift velocity $W^{*}$, whatever that
be (trajectories corresponding to $V < W^{*}$ from the gas on the left
will never again intersect the piston's trajectory).  Similarly, the
piston will encounter only those particles from
the gas on its right as have velocities {\it less} than $W^{*}$.  
Moreover, after each collision, the piston simply acquires the
velocity of the particle it collided with. 
Hence
the stationary velocity distribution must necessarily be a superposition
of the form
\begin{equation}
P_{\rm st}(V) \propto  n^{-}\phi^{-}(V) \,
\theta (V-W^{*}) + n^{+} \phi^{+} (V) 
\,\theta (W^{*} - V)
\label{Pstationary2}
\end{equation}
apart from a normalization factor.  The latter is easily seen to be just
$1/\Xi(W^{*})$ where the function 
$\Xi$ is as given by Eq. (\ref{chiwbar}).  Taking
the first moment of this equation, we find that consistency demands that
$W^\ast$ satisfy precisely the same implicit equation
(Eq. (\ref{Wbar1})) as that
written down above for the drift velocity $\overline{W}$.  In other words,
not only is the form in Eq. (\ref{Pstationary1}) 
for $P_{\rm st} (V)$ deducible on direct
physical grounds, but also the existence of an asymptotic drift velocity
$\overline{W}$ and the implicit equation satisfied by it.

\section{Exact expression for energy transfer}
\label{exactexpression}

To derive of an exact formula for the rate of transport of energy across
the piston, we first calculate the difference $\Delta E(t)$ between the
mean energy $E(t)$ of the gas on the left of the piston at time $t$ and
its initial value $E(0)$.

We have
\begin{equation}
E(0) = \textstyle{\frac{1}{2}}m \,\sum_{j=-N^{-}}^{-1} \langle V_{j}^{2}\rangle =
 \textstyle{\frac{1}{2}}m\,\sum_{j=-N^{-}}^{N^{+}} \big\langle V_{j}^{2} \,\theta (-X_{j})
\big\rangle\,,
\label{ezero}
\end{equation}
while
\begin{equation}
E(t) =  \textstyle{\frac{1}{2}}m \,\sum_{a} \,\sum_{j\neq a} 
\Big\langle V_{j}^{2} \,\theta (X_{a}(t) - X_{j}(t))\,\delta^{\rm Kr} 
[N^{-},\,\sum_{\ell \neq a} \theta (X_{a}(t) - X_{\ell}(t))]\Big\rangle\,. 
\label{etee}
\end{equation}
Here each sum runs over the entire set of particle labels
$\{-N^{-},\ldots,-1,0,1,\ldots,N^{+}\}$; the subscript $a$ is used to
distinguish the instantaneous variables of the piston from those of the
other particles; $\delta^{\rm Kr}[n,m]$ is the Kronecker delta $\delta_{nm}\,$;
and $\langle\ldots\rangle$ denotes an average over the already-specified
distributions of the initial conditions comprising 
the set $\{X_{j}\,,\,V_{j}\}$. 
Using the crucial
fact that
\begin{equation}
\sum_{\ell\neq a} \theta \big(X_{a}(t) - X_{\ell}(t)\big) = N^{-}
\label{constraint}
\end{equation}
for all $t \geq 0$, the gain in energy of the gas on the left at time $t$
can be written as
\begin{equation}
\Delta E(t) = 
\sum_{a} \sum_{j\neq a} \Big\langle
V_{j}^{2}\big[\theta \big(X_{a}(t) - X_{j}(t)\big) - \theta(-X_{j})\big]
\,\delta^{\rm Kr}\big[N^{-},\,\sum_{\ell\neq a} \theta \big(X_{a}(t) -
X_{\ell}(t)\big)\big]\Big\rangle. 
\label{deltaetee1}
\end{equation}
The representation $\delta_{nm} = (2\pi i)^{-1}\oint z^{n-m-1}\,dz$ enables us
write this as
\begin{eqnarray}
\Delta E(t)& = &\textstyle{\frac{1}{2}}m \displaystyle{\oint
\frac{dz}{2\pi iz}}
\,\sum_{a}\,\sum_{j\neq a}\Big\langle V_{j}^{2}\,\big[\theta (X_{a} +
  V_{a}\,t -
X_{j} - V_{j}\,t) - \theta(-X_{j})\big]\nonumber\\ 
&& \times \,z^{N^{-}} \prod_{\ell\neq a} \big[1 +
(z^{-1} -1) \,\theta(X_{a} + V_{a}\,t - X_{\ell} - V_{\ell} \,t)\big]
\Big\rangle\,,
\label{deltaetee2}
\end{eqnarray}
where the contour encloses the origin. We break up
$\Delta E(t)$ as
\begin{equation}
\Delta E(t) = \Delta E^{0}(t) + \Delta E^{-}(t) + \Delta E^{+}(t)\,,
\label{deltaetee3}
\end{equation}
corresponding to the three distinct
contributions to $\Delta E(t)$ coming, respectively, from the cases $a =
0$, $-N^{-} \leq a \leq -1$ and $1 \leq a \leq N^{+}$.
The calculation of these quantities is broadly similar to, but a little
more intricate than, that involved \cite{lebo,bala1} in 
evaluating quantities
like the one-particle distribution function 
$P(X,V,t \arrowvert X_{0},V_{0})$ of the
piston.  An outline of the main steps leading to the expression for
$\Delta E(t)$ is given in the Appendix.  The exact result for $\Delta
E(t)$ is given by Eqs. (\ref{deltaezero2})-(\ref{deltaeplus2}).

\section{Formula for the stationary energy flux}
\label{fluxformula}

To find the asymptotic or steady-state rate of transport of energy across
the piston, we pass to the long-time limit of the expression for $\Delta
E(t)$.  Using the asymptotic behavior $I_{r}(z) \sim e^{z}/(2\pi z)^{1/2}$ as
$|z| \rightarrow \infty$, it is evident from Eq. (\ref{deltaezero2}) 
that $\Delta E^{0}(t)$ 
decays exponentially to zero as $t \rightarrow \infty$, unless it so
happens that $n^{-}\,\alpha(0) = n^{+}\,\beta(0)$, 
i.e., the asymptotic drift
velocity of the piston is zero.  In that case $\Delta E^{0} (t) \sim
t^{1/2}$ at long times; but this again implies a rate of transport that
vanishes (like $t^{-1/2}$) as $t \rightarrow \infty$.  Therefore $\Delta
E^{0}(t)$ may be dropped from further consideration.

Turning to Eqs.
(\ref{deltaeminus2}) and (\ref{deltaeplus2}) for 
$\Delta E^{-}(t)$ and $\Delta E^{+}(t)$, we observe that the asymptotic
behavior of the modified Bessel functions leads to the occurrence of a
factor $\exp \,\left\{-t\left[\big(n^{-}\alpha(w)\big)^{1/2} 
- \big(n^{+}\beta (w)\big)^{1/2}\right]^2\right\}$ in the
integration over $w$ (which runs from $-\infty$ to $+\infty$). Since the
exponent has a unique zero at precisely the asymptotic drift velocity
$\overline{W}$ as defined in Eq. (\ref{Wbar1}), the long-time behavior of
$\Delta E^{\pm}(t)$ can be deduced by a standard Gaussian approximation
about $w = \overline{W}$.  The leading asymptotic behavior of each of
these two contributions is then seen to be $\sim t$, implying the
existence of a finite, non-vanishing stationary rate of energy transfer
$\lim_{t \rightarrow \infty} d \Delta E(t)/dt$, that we denote by
$\Delta \dot{E}_{\rm st}$.  Carrying out the algebra required, our final
result for this quantity is remarkably simple (for reasons to be explained
shortly). We find
\begin{equation}
\Delta \dot{E}_{\rm st} = \textstyle{\frac{1}{2}}m \Big\{n^{+}
\int_{-\infty}^{\overline{W}} dU \, U^{2}(\overline{W} - U) \,\phi^{+}(U)
- n^{-} \int_{\overline{W}}^{\infty} dU\, U^{2} (U - \overline{W})\, 
\phi^{-}(U)\Big\}.
\label{flux1}
\end{equation}
For ready reference, we recall that the asymptotic drift velocity
$\overline{W}$ of the piston is given by the implicit equation
(\ref{Wbar1}), which may be written in the alternative form
\begin{equation}
\overline{W} = \frac{n^{-}\int_{\overline{W}}^{\infty} \,dU\,
U\,\phi^{-}(U) + n^{+} \int_{-\infty}^{\overline{W}}\, dU\, U\, \phi^{+}
(U)}{n^{-} \int_{\overline{W}}^{\infty} \,dU \,\phi^{-}(U) + n^{+}
\int_{-\infty}^{\overline{W}} \,dU \,\phi^{+} (U)}\,.
\label{Wbar2}
\end{equation}
We remark that the result in Eq. (\ref{flux1}) is 
valid for arbitrary
initial velocity distributions $\phi^{\pm}(V)$, and not just for
Maxwellian $\phi^{\pm}$ (see below). It must also be noted that 
the existence of the {\it third}
moment of the speed, $\langle |V|^3\rangle^{\pm}$, is necessary in order
that $\Delta \dot{E}_{\rm st}$ be finite.
The exact result in Eq. (\ref{flux1}) has several other 
noteworthy features, that we take up in turn.\\

\noindent
$(i)$\,{\it Physical interpretation}:\,The formula in Eq.
(\ref{flux1}) can be given a direct physical 
interpretation, extending the heuristic 
argument described in Sec. \ref{notation}
for $P_{\rm st} (V)$ and $\overline{W}$: once
again, the fact that the particles merely exchange velocities upon
colliding implies that the mean stationary rate of energy transfer across
the piston can be constructed by an energy balance argument. If the piston
has a stationary drift velocity $\overline{W}$, the trajectories belonging
originally to the gas on its right that collide with it are those
corresponding to velocities from $-\infty$ to $\overline{W}$.  The energy
carried by (a particle on) each such trajectory is $\frac{1}{2}
mU^{2}$.  The rate of
collisions of such trajectories with that of the piston is
$n^{+}\,(\overline{W}-U)$, the second factor being the relative velocity
between the two. Therefore
\begin{equation}
\frac{m n^{+}}{2} \int_{-\infty}^{\overline{W}} dU \,
U^{2} \,(\overline{W}-U)\,\phi^{+} (U)
\label{rightrate}
\end{equation}
is the mean rate at which energy is transferred to the piston by the gas
on its right, in the stationary state. Exactly the same argument shows
that
\begin{equation}
\frac{m n^{-}}{2} \int_{\overline{W}}^{\infty} dU\, U^{2} \,
(U - \overline{W})\,\phi^{-}(U)
\label{leftrate}
\end{equation}
is the mean rate at which the gas on the left transfers energy to the
piston. The difference between the two is precisely the formula of Eq.
(\ref{flux1}) for the mean stationary rate of transfer 
of energy from the
right to the left across the piston. Our rigorous derivation serves to
corroborate this physical argument, in addition to providing an exact
result for the time-dependent transients as well.

\noindent
$(ii)$\,{\it Dichotomous and three-valued velocity distributions}:\,
The dichotomous velocity distribution
\begin{equation}
\phi^{+}(V) = \phi^{-}(V) = \phi(V) = 
\textstyle{\frac{1}{2}}\,\big[\delta (V + c) +
\delta(V-c)\big]
\label{dichdist}
\end{equation}
yields, as always, a simple and tractable special case that serves as a
useful check on the calculations. We have, in this case,
$\overline{W} = c \,(n^{-} - n^{+})/(n^{-} + n^{+})$.
Evaluating the various quantities appearing in 
Eq. (\ref{flux1}), we find
that $\Delta \dot{E}_{\rm st}$ vanishes identically for the dichotomous
distribution above.  However, valuable insight into the structure of the
result for $\Delta \dot{E}_{\rm st}$ is provided by a slight generalization of
the dichotomous distribution of Eq. (\ref{dichdist}) to the distribution
\begin{equation}
\phi^{+}(V) = \phi^{-}(V) = \phi(V) = 
\mu \,\delta(V) + \textstyle{\frac{1}{2}} 
(1-\mu)\,\left[\delta(V+c) + \delta(V-c)\right]\,, 
\label{trichdist}
\end{equation}
where $0 < \mu < 1$.  It has been shown in Ref. \cite{bala1} 
that the introduction of
a non-vanishing probability mass $\mu$ at $V = 0$ significantly alters the
long-time properties of the homogeneous system.  For instance, the
velocity autocorrelation function acquires a $t^{-3/2}$ tail, in
contrast to its exponential decay in the case of a dichotomous
$\phi(V)$. In the
present context, too, a non-vanishing value of $\mu$ leads to a strikingly
different result for $\Delta \dot{E}_{\rm st}$.  We have in this instance
\begin{equation}
\overline{W} = \frac{c(1-\mu)(n^{-} - n^{+})}{(1-\mu) n_{>} + (1 +
\mu)n_{<}}\,,
\label{trichwbar}
\end{equation}
where $n_{>} = \max \, (n^{-},n^{+})$ and $n_{<} = \min\, (n^{-}, n^{+})$.  
Computing the various quantities occurring in Eq. (\ref{flux1}), 
we arrive at the result
\begin{equation}
\Delta \dot{E}_{\rm st} = 
\frac{m c^{3}}{2} \frac{\mu\,(1-\mu)\,n^{-}\,(n^{+} -
n^{-})}{(1-\mu)\,n_{>} + (1+\mu)\,n_{<}}\,. 
\label{trichflux}
\end{equation}
Thus $\Delta \dot{E}_{\rm st}$ vanishes identically if $n^{+} = n^{-} = n$ 
(which, together with $\phi^{+} = \phi^{-}$ as already imposed by 
Eq. (\ref{trichdist}),
implies a homogeneous system), as it must in the homogeneous system.  
Similarly, $\Delta \dot{E}_{\rm st}$ vanishes when $\mu = 0$ (the case of a
dichotomous $\phi^{\pm}(V)$), or when $\mu = 1$ (the trivial case of no
motion at all). Comparing the expressions in Eqs. (\ref{trichwbar}) and
(\ref{trichflux}), we observe that $\Delta \dot{E}_{\rm st}$ can be written 
in this case in the revealing form
\begin{equation}
\Delta \dot{E}_{\rm st} = 
\textstyle{\frac{1}{2}}\,m c^{2} \,\mu \,n^{-} (-\overline{W}).
\label{trichrelation}
\end{equation}
(Recall that we have defined $\Delta \dot{E}$ as the rate of transfer of
energy to the gas on the {\em left} of the piston, and that $\overline{W}$
is negative if $n^{+} > n^{-}$, i.e., if the piston drifts to the left.)

\noindent
$(iii)$\,{\it The case of Maxwellian distributions}:\, 
The situation that is of direct physical interest is of course
that of Maxwellian velocity distributions $\phi^{\pm}(V)$ characterized by
temperatures $T^{\pm}$.  For arbitrary densities $n^{+}$ and $n^{-}$, the
asymptotic drift velocity $\overline{W}$ is still given by the solution of
a transcendental equation, which reads in this case
\begin{equation}
\overline{W} = \frac{n^{-}\,(2k_{B}T^{-}/m\pi)^{1/2}\,e^{
-m\overline{W}^{2}/2k_{B}T^{-}} - n^{+} \,(2k_{B} T^{+}/m\pi)^{1/2}\,
e^{-m\overline{W}^{2}/2k_{B}T^{+}}}
{n^{-}\,\big[1-\mbox{erf}\,\,
(m\overline{W}^{2}/2k_{B}T^{-})^{1/2}\big] + 
n^{+}\,\big[1+\mbox{erf}\,\,
(m\overline{W}^{2}/2k_{B}T^{+})^{1/2}\big]}\,.
\label{wbarmaxwellian}
\end{equation}
In terms of $n^{\pm}$, $T^{\pm}$ and $\overline{W}$ as given above, we get
\begin{eqnarray}
\Delta \dot{E}_{\rm st}&=& (2\pi m)^{-1/2}\left\{n^{+}(k_{B}T^{+})^{3/2} 
e^{-m\overline{W}^{2}/2k_{B}T^{+}} - n^{-} (k_{B}T^{-})^{3/2} 
e^{-m\overline{W}^{2}/2k_{B}T^{-}}\right\}\nonumber\\
&& + \textstyle{\frac{1}{4}}\overline{W}\left\{n^{+}
(m\overline{W}^{2} + k_{B}T^{+})\big[1+\mbox{erf}\,\,
(m\overline{W}^{2}/2k_{B}T^{+})^{1/2}\big]\right.\nonumber\\
&& +\left. n^{-} (m\overline{W}^{2} +
k_{B}T^{-})\big[1-\mbox{erf}\,\, (m\overline{W}^{2}/2k_{B}T^{-})^{1/2}
\big]\right\}\,.
\label{fluxmax}
\end{eqnarray}
The nonlinear dependence of $\Delta \dot{E}_{\rm st}$ on $\overline{W}$ is
explicit in this formula.  We shall comment further on this subsequently.  
It is also clear that $\Delta \dot{E}_{\rm st}$ is {\em not} proportional to
the temperature difference $(T^{+}-T^{-})$, though one might perhaps naively
expect such a proportionality based on an incorrect identification of
$\Delta \dot{E}_{\rm st}$ with the heat flux, together with an application of
the Fourier law to the system under discussion.

\noindent
$(iv)$\,{\it Mechanical work and heat contributions}:\,
This brings us,
finally, to a very important point.  The general expression obtained 
in Eq. (\ref{flux1}) for $\Delta \dot{E}_{\rm st}$ incorporates the
effects (on the energy transfer rate) of both the {\it drift} of the
piston and its {\it diffusive} motion (or fluctuations about its mean
position, namely, about $\overline{W} t$). In broad terms, one might
regard the respective contributions as the rate of mechanical work 
 done upon the gas on the left of the piston
($\overline{W}$ being the rate of compression or expansion, depending on
whether $\overline{W} < 0$ or $\overline{W} > 0$), and the rate at which
its entropy changes.  However, these contributions are intricately mixed
up in the formula for $\Delta \dot{E}_{\rm st}$. 
Moreover, the dependence of
$\Delta \dot{E}_{\rm st}$ upon $\overline{W}$ is in general highly nonlinear.  
Disentangling these contributions will enable us, in principle, to isolate
the parts of $\Delta \dot{E}_{\rm st}$ that may be identified with the ``heat
flux'' $\Delta \dot{Q}_{\rm st}$ and the ``rate of mechanical work'',
respectively.  In some cases, the former may vanish altogether --- as in
the example of the distribution considered in Eq. (\ref{trichdist}), for which
Eq.  (\ref{trichrelation}) shows clearly 
that $\Delta \dot{E}_{\rm st}$ arises entirely
from the drift of the piston.  In general, however, such a clear
separation does not occur in the system under study. It is also important
to bear in mind the fact that a non-vanishing drift velocity
$\overline{W}$ is itself a consequence of the statistics of collisions in
the system under consideration \cite{pias}. 

\section{Stationary heat flux}
\label{statflux}

A direct way to isolate and examine the ``heat flux'' is to impose the
condition of zero drift $(\overline{W} = 0)$ by adjusting, for instance,
the value of the ratio $n^{-}/n^{+}$ of the densities of the two gases.  
As we
shall now see, this leads to considerable simplification in the formula
for $\Delta \dot{E}_{\rm st}$, which we are now justified in re-labeling
as $\Delta
\dot{Q}_{\rm st}$.

As emphasized more than once, the
asymptotic drift velocity $\overline{W}$ of the piston vanishes if the
densities $n^{\pm}$ and initial velocity distributions $\phi^{\pm}$ are
related by the condition 
$n^{-} \alpha(0) = n^{+} \beta (0)$, or $n^{-}\langle
|V|\rangle^{-} = n^{+} \langle |V|\rangle^{+},$ i.e., the mean rates at which
the piston suffers collisions from the gases on either side of it are
equal.  The motion of the piston is then purely diffusive in the long-time
limit, with a variance $\langle X^{2} (t)\rangle$ that tends asymptotically
to $2Dt$, where the diffusion coefficient is given by 
\cite{bala1,bala2}
\begin{equation}
D = 2n^{-} \langle |V|\rangle^{-}/(n^{-} + n^{+})^{2}\,. 
\label{diffusioncoeff}
\end{equation}
We note that the homogeneous system (defined by $n^{-} = n^{+}, \phi^{-} =
\phi^{+}$) automatically has $\overline{W} = 0$. The converse is not
necessarily true, of course: $\overline{W} = 0$ does not necessarily imply
a homogeneous system. Recall also that we have already considered
situations in which 
$\phi^{+}(V) = \phi^{-}(V) = \phi(V)$, but $n^{+} \neq n^{-}$;
we then have an inhomogeneous system in which, moreover, $\overline{W}
\neq 0$.

Setting $\overline{W} = 0$ in Eq. (\ref{flux1}), we find
that the stationary energy flux in the absence of drift is simply
\begin{equation}
\Delta \dot{Q}_{\rm st} = \textstyle{\frac{1}{4}}\,m 
\big[n^{+} \langle |V|^3\rangle^{+} - n^{-} 
\langle |V|^3\rangle^{-}\big]\,,
\label{heatflux}
\end{equation}
in terms of the third moments
\begin{equation}
\langle |V|^3\rangle^{\pm} = 2\int_{0}^{\infty} dU\, U^{3}
\,\phi^{\pm}(U)
\label{thirdmoment}
\end{equation}
of the particle speed in the two gases.  For the homogeneous system, of
course, $\Delta \dot{Q}_{\rm st}$ vanishes identically, as it must.

Turning again to the case of Maxwellian distributions $\phi^{\pm}(V)$ at
temperatures $T^{\pm}$, the drift velocity $\overline{W}$ vanishes, as is
well known, if $n^{-}\sqrt{T^{-}} = n^{+} \sqrt{T^{+}}$.  
Equation (\ref{fluxmax}) then reduces to
\begin{equation}
\Delta \dot{Q}_{\rm st} = n^{+} \left(\frac{k_{B}T^{+}}{2\pi m}\right)^{1/2}
k_{B}\,(T^{+} - T^{-}). 
\label{heatfluxmax}
\end{equation}
We see that the heat flux is indeed given by the Fourier law in this case,
with a ``coefficient of thermal conductivity'' that is proportional to the
square root of the temperature.

It is revealing and instructive to
pause at this stage to compare the result in 
Eq. (\ref{heatfluxmax}) with that for
the heat flux between two reservoirs held at different temperatures and
densities, coupled by virtue of their sharing 
a common piston \cite{kest}. When
the mass of the latter is equal to that of the gas particles, the
Boltzmann equation can be solved exactly for the stationary distribution;
concomitantly, the heat flux conveyed via the shared piston from one
reservoir to the other can also be calculated. While the exact expression
for this quantity (see \cite{kest}) is slightly more complicated than that of
Eq. (\ref{heatfluxmax}) above, 
it reduces to the latter result to leading order in
the temperature difference $(T^{+} - T^{-})$, apart from an extra
numerical factor $\pi/\sqrt{2}\,$. This lends additional support to our
identification of the portion of the energy flux that corresponds to the
heat flux. The fact that the heat flux is about twice as large in the case
of the shared piston is readily understood by recalling that the piston
now suffers collisions with the gas particles belonging to both
reservoirs, so that the effective rate of collisions is roughly twice as
large. For a detailed discussion of the relevance of the problem of the
shared piston to the general questions addressed here, we 
refer to \cite{kest}.

Finally, let us return to the full expression for the stationary energy
flux $\Delta \dot{E}_{\rm st}$ in Eq. (\ref{flux1}), and analyze its dependence
on the drift velocity $\overline{W}$.  For sufficiently small
$\overline{W}$, we may expand $\Delta \dot{E}_{\rm st}$ in powers of
$\overline{W}$, taking care to incorporate the fact that $\overline{W} =
0$ imposes the condition $n^{-} \langle |V|\rangle^{-} = n^{+} \langle
|V|\rangle^{+}$ on the parameters occurring in the coefficients of the
expansion. (This has been done in the Maxwellian
case, Eq. (\ref{heatfluxmax}) above, to
reduce $n^{+}(T^{+})^{3/2} -
n^{-}(T^{-})^{3/2}$ to $n^{+} (T^{+})^{1/2} (T^{+}-T^{-})$.)  
For distributions $\phi^{\pm}(V)$ that have derivatives of all orders at
$V = 0$, the formal expansion in powers of $\overline{W}$ is given by
\begin{equation}
\Delta \dot{E}_{\rm st} = \Delta \dot{Q}_{\rm st} + \frac{m \overline{W}}{4}
\big[n^{+}
\langle V^{2}\rangle^{+} + n^{-} \langle V^{2}\rangle^{-}\big] 
+ \frac{m \overline{W}^{\,\,4}}{2} \sum_{k=0}^{\infty}
\frac{\overline{W}^{\,\,k} \psi^{(k)}(0)}{(k+3)(k+4)k!}\,\,, 
\label{flux2}
\end{equation}
where 
\begin{equation}
\psi(V) = n^{+} \phi^{+} (V) - n^{-} \phi^{-}(V)\,, 
\label{psi}
\end{equation}
$\psi^{(k)}$ denotes its
$k^{\rm th}$ derivative, and $\Delta \dot{Q}_{\rm st}$ 
is given by Eq. (\ref{heatflux}).  
This representation isolates the contribution to $\Delta \dot{E}_{\rm st}$
owing to the systematic drift of the piston from that arising from the
fluctuations about its mean position.  We observe that the part that is
nonlinear in $\overline{W}$ is ${\cal O}(\overline{W}^4)$.  Considering the
Maxwellian case once again, we find
\begin{equation}
\Delta \dot{E}_{\rm st} \approx 
\left(\frac{k_{B}T^{+}}{2\pi m}\right)^{1/2} n^{+}
\,k_{B} \,(T^{+}-T^{-}) + \left(\frac{k_{B} T^{+}T^{-}}{8\pi m}\right)^{1/2}
k_{B}\,\big(n^{-}\sqrt{T^{-}} - n^{+}\sqrt{T^{+}}\big) 
\label{5.8}
\end{equation}
correct to first order in the difference $\big(n^{-}\sqrt{T^{-}} -
n^{+}\sqrt{T^{+}}\big)$, the next term being of fourth 
order in this quantity. In
contrast to the relatively simple expression to which $\Delta
\dot{E}_{\rm st}$ reduces when $n^{-}\sqrt{T^{-}} = n^{+}\sqrt{T^{+}}$ 
(Eq. (\ref{heatfluxmax})), no significant 
simplification of the general expression in 
Eq. (\ref{fluxmax}) occurs when
$n^{-}T^{-} = n^{+}T^{+}$ (equal {\em pressures} 
on either side of the piston).  
Some simplification does occur, however, when the {\em temperatures}
on the two sides are equal, $T^{+} = T^{-} = T$ (but $n^{+} \neq n^{-}$).  
We find
that the drift velocity is now given by the implicit equation
\begin{equation}
\overline{W} = \left(\frac{n^{-}-n^{+}}{n^{-} + n^{+}}\right)
\left[\left(\frac{2k_{B}T}{m\pi}\right)^{1/2} 
e^{-m\overline{W}^{2}/2k_{B}T} + \overline{W} \,\mbox{erf}\,\,
(m\overline{W}^{2}/2k_{B}T)^{1/2}\right].
\label{maxwbareqtemp}
\end{equation}
The corresponding stationary energy flux is found to be
\begin{equation}
\Delta \dot{E}_{\rm st} = (n^{-}-n^{+})
\left(\frac{mk_{B}T}{8\pi}\right)^{1/2}\left(\overline{W}^{2} -
\frac{k_{B}T}{m}\right)\,
e^{-m\overline{W}^{2}/2k_{B}T}\,.
\label{maxfluxeqtemp1}
\end{equation}
Written in terms of the moments of the (Maxwellian) velocity distribution,
this is just
\begin{equation}
\Delta \dot{E}_{\rm st} = \textstyle{\frac{1}{4}}
\,m\,\left(\overline{W} - \langle V^{2}\rangle\right) 
\,(n^{-}-n^{+})\, \langle \,|V|\,\rangle \,
e^{-m\overline{W}^{2}/2k_{B}T}\,.
\label{maxfluxeqtemp2}
\end{equation}
Finally, if the two {\it densities} are equal $(n^{+} = n^{-} = n$, 
but $T^{+}
\neq T^{-})$, we find that $\overline{W}$ becomes independent of $n$, and
$\Delta \dot{E}_{\rm st}$ is directly proportional to $n$.

\section{Concluding remarks}
\label{conclusion}

We have already commented at the appropriate junctures on 
the special and interesting 
features of the structure of our analytical result for the stationary rate of
energy transfer, as given by Eqs. (\ref{flux1}) and (\ref{flux2}). 
We conclude with a comment on
the question of the finiteness or otherwise of the heat conductivity
of the system under study\cite{dhar,lepr}.
It may be argued that the coefficient of heat conductivity
of our system is essentially infinite: The contention is that in the expression
for the heat flux, the conductivity is the coefficient of the gradient
of the temperature, and the latter is $\propto (T^{+}-T^{-})/L$. This
gradient
tends to zero in the thermodynamic limit, although a finite
energy flux persists in the long time limit. Hence the coefficient multiplying
the gradient has to diverge. 

While we agree with this argument, we point out that the system can 
be viewed in a different way: the gases on the left and right of the piston
are heat reservoirs which have a single microscopic degree of contact,
namely, the tagged particle we have termed the piston. There is no relevant
length scale in the problem, so the ``heat flux'' is expected to appear
solely as a result of this thermal contact. The fact that the Boltzmann
calculation gives a comparable value for the conductivity supports the
meaningfulness of our result. Furthermore, one
can construct a physical system that will display exactly the behavior
predicted by our model calculation. Consider a two- or three-dimensional 
cylinder of axial length $2L$ with a central piston that 
can move without friction
along the axis of the cylinder. The compartments to its left and right
are filled with gases in equilibrium, at respective
densities and temperatures $n^{\pm}, T^{\pm}$. In the limit of low
densities and $L \rightarrow \infty$, the dynamics of the piston will
be exactly as described by the one-dimensional model: in the physical
set-up, the particles only interact with the piston, while in our 
one-dimensional model, they exchange velocities upon collision, which
is effectively tantamount to their passing through each other 
without interaction.

\section*{Acknowledgment}
VB acknowledges the warm
hospitality of the Limburgs Universitair Centrum during his visit
there, when part of this work was carried out.

\newpage

\appendix*
\section{Calculation of $\Delta E(t)$} 

As stated in the main text, we write Eq. (\ref{deltaetee2}) for 
$\Delta E(t)$ as
the sum $\Delta E^{0}(t) + \Delta E^{-}(t) + \Delta E^{+}(t)$
of contributions coming, respectively, from the possibilities that 
at time $t$, the piston is (i) on its original trajectory, so that $a =0$, or
(ii) on a trajectory belonging to the gas on its left, so that 
$-N^{-} \leq a \leq -1$, or (iii) on a trajectory belonging to the gas on its
right, so that  $1 \leq a
\leq N^{+}$. Consider $\Delta E^{0}$ first.  We have in this case 
$X_{0} = 0$, $V_{0} = 0$. We find
\begin{eqnarray}
\Delta E^{0} (t) &=& m\oint \frac{dz}{4\pi iz} \sum_{j\neq
0} \big\langle V_{j}^{2} \left[\theta (-X_{j} - V_{j}\,t) 
- \theta (-X_{j})\right]
\big\rangle \nonumber  \\
&& \times
\prod_{\ell -} \left[z+ (1-z) \big\langle \theta (-X_{\ell} - V_{\ell}\,t)
\big\rangle\right] \prod_{\ell'+} \left[1 +(z^{-1}-1) 
\big\langle \theta (-X_{\ell'} 
- V_{\ell '} t)\big\rangle\right],
\label{deltaezero1}
\end{eqnarray}
where the symbols $\ell -$ and $\ell' +$ are used to denote the 
fact that $\ell$ and $\ell'$ run
over $-N^{-} \leq \ell \leq -1$ and $1 \leq \ell' \leq N^{+}$, respectively, in
the products concerned.  We shall use the convenient notation $\langle
\ldots \big\rangle_{j}^{\pm}$ for the corresponding averages. 
The evaluation of
the contributions $\Delta E^{\pm}(t)$ is somewhat more involved.  We get
\begin{eqnarray}
\Delta E^{-}(t) &=& m\oint \frac{dz}{4\pi iz} N^{-} 
\left\langle \left\{(N^{-}-1) 
\big\langle V_{j}^{2}\,[\theta(X_{a} + V_{a}\,t - X_{j} - V_{j}\,t) -
\theta(-X_{j})]\big\rangle_{j}^{-} \phantom{A^{n^{+}}}\right.\right. 
\nonumber \\
&& \left. +\,N^{+}\,\big\langle V_{j}^{2} \,[\theta(X_{a} + V_{a}\,t
- X_{j} - V_{j}\,t) - \theta(-X_{j})]\big\rangle_{j}^{+}
\phantom{A^{n^{+}}}\hspace{-0.8cm}
\right\} \nonumber \\
&&\times \,[z + (1-z)\big\langle
\theta(X_{a} + V_{a}\,t - X_{\ell}-V_{\ell}
t)\big\rangle_{\ell}^{-}]^{N^{-}-1}\;
[z + (1-z)
\,\theta(X_{a} + V_{a} t)] \nonumber \\
&& \left.\,\times [1 + (z^{-1} -1) \big\langle \theta(X_{a} + V_{a}\,t -
X_{\ell'}-V_{\ell'}t)\big\rangle_{\ell'}^{+}]^{N^{+}}\right\rangle_{a}^{-},
\label{deltaeminus1}
\end{eqnarray}
while
\begin{eqnarray}
\Delta E^{+}(t)&=& m
\oint \frac{dz}{4\pi iz} N^{+}\left\langle
\left\{N^{-}\big\langle V_{j}^{2} [\theta (X_{a} + V_{a}\,t - X_{j} -V_{j}\,t) -
\theta(-X_{j})]\big\rangle_{j}^{-}\phantom{A^{n^{+}}}\right.\right. 
\nonumber \\
&&\left. + \,(N^{+} -1) \big\langle V_{j}^{2} \,[\theta(X_{a} + V_{a}\,t -
X_{j} - V_{j}\,t) - \theta(-X_{j})]\big\rangle_{j}^{+}\phantom{A^{n^{+}}}
\hspace{-0.8cm}\right\} \nonumber\\
&& \times \,[z+(1-z) \big\langle \theta (X_{a}
+ V_{a}\,t - X_{\ell} - V_{\ell} t)\big\rangle_{\ell}^{-}]^{N^{-}} \,[1 +
(z^{-1}-1)\,\theta(X_{a}+V_{a}\,t)]\nonumber\\
&&\left.\times \,[1 + (z^{-1}-1) \big\langle \theta (X_{a} + V_{a}\,t
- X_{\ell'} - V_{\ell'}t)
\big\rangle_{\ell'}^{+}]^{N^{+}-1}\right\rangle_{a}^{+}.
\label{deltaeplus1}
\end{eqnarray}
Evaluating the averages required, we get, for instance,
\begin{eqnarray}
&\big\langle V_{j}^{2}[\theta (-X_{j}-V_{j}\,t) - 
\theta (-X_{j})]\big\rangle_{j}^{-} 
= -\int_{L/t}^{\infty} dU\, U^{2}\, \phi^{-}(U) - (t/L) 
\int_{0}^{L/t} dU \,U^{3} \,\phi^{-}(U)\,, \nonumber \\
&\big\langle V_{j}^{2} \,
[\theta(-X_{j}-V_{j}\,t) - \theta(-X_{j})]\big\rangle_{j}^{+} = 
\int_{-\infty}^{-L/t} dU\, U^{2} \,\phi^{+}(U) - 
(t/L)\int_{-L/t}^{0} dU \,U^{3}\, \phi^{+} (U)\,, \nonumber \\
&\big\langle \theta(-X_{\ell} - V_{\ell} t)\big\rangle_{\ell}^{-} 
= \int_{-\infty}^{L/t} dU \,\phi^{-}(U) - (t/L) \int_{0}^{L/t} dU\, U
\,\phi^{-}(U)\,,
\nonumber \\
& \big\langle \theta (-X_{\ell'} - V_{\ell'}t)\big\rangle_{\ell'}^{+} 
= \int_{-\infty}^{-L/t} dU \,\phi^{+} (U) 
- (t/L) \int_{-L/t}^{0} dU\, U \,\phi^{+} (U)\,.
\label{averages}
\end{eqnarray}
The other averages are similarly calculated. Inserting all these results
in Eqs. (\ref{deltaezero1}) - (\ref{deltaeplus1}), 
we pass to the thermodynamic limit $N^{\pm}
\rightarrow\infty$, $L \rightarrow \infty$, such 
that $\lim N^{\pm}/L = n^{\pm}$.  Using
the well-known expression for the generating function of the modified
Bessel function $I_{r}(z)$, the final expressions are as follows:

Recall the definitions of the rates
$\alpha (w)$ and $\beta (w)$ in 
 Eq. (\ref{alphabeta1}), and let
\begin{equation}
\lambda^{-}(w \,;\, n^{-}) 
= n^{-}\,\alpha (w)\,,\,\, \lambda^{+} (w \,;\, n^{+}) 
= n^{+} \,\beta (w).
\label{lambdapm}
\end{equation}
Define the effective rates 
\begin{equation}
\lambda (w\,; \,n^{-}, n^{+}) = (\lambda^{-}\lambda^{+})^{1/2}\,
, \,\,
\Lambda(w\,;\,n^{-},n^{+}) = \lambda^{-} + \lambda^{+}\,.
\label{Lambda}
\end{equation}
Further, let
\begin{equation}
F(w\,;\,n^{-}, n^{+}) = n^{-}\int_{w}^{\infty} dU \, U^{2} (w-U) 
\,\phi^{-}(U) + n^{+} \int_{-\infty}^{w} dU\, U^{2} (w - U) 
\,\phi^{+}(U)\,.
\label{F}
\end{equation}
Then, suppressing the $n^{\pm}$ dependence in $\lambda$ and $\Lambda$ for
notational simplicity, we find
\begin{equation}
\Delta E^{0} (t) = \textstyle{\frac{1}{2}}\,m \,t \,e^{-\Lambda(0)t} \,I_{0} 
\big(2\lambda (0)t\big)\, F(0\,;\,n^{-}, n^{+})\,, 
\label{deltaezero2}
\end{equation}
\begin{eqnarray}
\Delta E^{-} (t)& =& \textstyle{\frac{1}{2}}\,mn^{-}t^{2} 
\int_{-\infty}^{\infty} 
dw\, e^{-\Lambda (w)t} 
\,\arrowvert\alpha'(w)\arrowvert\, F(w\,;\,n^{-},n^{+})\nonumber \\
&&\times \left\{\theta(w)\,I_{0} \big(2\lambda (w)t\big) + \theta (-w) 
\,[\lambda^{+} (w)/\lambda^{-} (w)]^{1/2}
\,I_{1} \big(2\lambda (w)t\big)\right\}\,,
\label{deltaeminus2}
\end{eqnarray}
\begin{eqnarray}
\Delta E^{+}(t)&=& \textstyle{\frac{1}{2}}\,mn^{+}t^{2} 
\int_{-\infty}^{\infty} 
dw\,e^{-\Lambda (w)t}\,\beta' (w)\, F(w\,;\,n^{-},n^{+})\nonumber\\
&&\times \left\{\theta(-w) \,I_{0} \big(2\lambda (w)t\big) 
+ \theta (w) \,[\lambda^{-} (w)/\lambda^{+} (w)]^{1/2} \,I_{1} 
\big(2\lambda (w)t\big)\right\}\,.
\label{deltaeplus2}
\end{eqnarray}
$\Delta E(t)$ is the sum of the right-hand 
sides of Eqs. (\ref{deltaezero2})-(\ref{deltaeplus2}).


\begin{thebibliography}{99}

\bibitem{dhar}
A fairly extensive literature exists on the subject. See, for
instance,
A. Dhar, Phys. Rev. Letters {\bf 86}, 3554 (2001); P. Grassberger,
W. Nadler, and L. Yang, Phys. Rev. Letters {\bf 89}, 180601 (2002); 
O. Narayan and S. Ramaswamy, Phys. Rev. Letters {\bf 89}, 200601
(2002); G. Casati and T. Prosen, Phys. Rev. E {\bf 67}, 015203(R)
(2003); S. Lepri, R. Livi, and A. Politi, Phys. Rev. E {\bf 68},
067102 (2003); C. Gruber and A. Lesne, {\tt cond-mat/0410413}; 
and references therein. 

\bibitem{lepr}
For a recent review and further references to the literature, see 
S. Lepri, R. Livi, and A. Politi, Phys. Rep. {\bf 377}, 1 (2003).

\bibitem{jeps} D. W. Jepsen, J. Math. Phys. {\bf 23}, 405 (1965).

\bibitem{lieb}
 E. Lieb, Physica A  {\bf 263}, 491 (1999); C. Gruber, Eur.
J. Phys. {\bf 20}, 259 (1999); E. Kestemont, C. Van den Broeck,
 and M. Malek Mansour, Europhys. Lett. {\bf 49}, 143 (2000);
N. I. Chernov, J. L. Lebowitz, and Ya. G. Sinai, Russ.
Math. Surv. {\bf 57}, 1045 (2002).

\bibitem{lebo} J. L. Lebowitz and J. K. Percus, Phys. Rev. 
{\bf 155}, 122 (1967).

\bibitem{pias} J. Piasecki, J. Stat. Phys. {\bf 104}, 1145 (2001).

\bibitem{bala1} V. Balakrishnan, I. Bena, and C. Van den Broeck, 
Phys. Rev. E {\bf 65}, 031102 (2002).

\bibitem{bala2} A factor of $4$ has been left out inadvertently 
in the numerator of 
the expression for the diffusion coefficient $D$ in Eq. (30) 
of Ref. \cite{bala1}.

\bibitem{kest} E. Kestemont {\it et al.}, in Ref. 
\cite{lieb}; C. Van den Broeck, E. Kestemont, 
and M. Malek Mansour, Europhys. Lett. {\bf 56}, 771 (2001).

\end{thebibliography}
\end{document}